\documentclass[reprint, twocolumn, superscriptaddress, aps, jcp]{revtex4-1}

\usepackage[
  unicode,
  pdfusetitle,    
  pdfcreator={},
  pdfproducer={}, 
]{hyperref}

\usepackage{mathtools}
\DeclareMathOperator\erf{erf}
\DeclareMathOperator\sgn{sgn}
\let\theta\vartheta
\newcommand*{\dd}{\mathrm{d}}
\newcommand*{\pp}{\mathrm{p}}
\newcommand*{\bu}{\textbf{u}}
\newcommand*{\bq}{\textbf{q}}

\begin{document}

\title{Continuum Percolation of Polydisperse Rods in Quadrupole Fields:\\ Theory and Simulations}
\author{Shari P. Finner}
\email{s.p.finner@tue.nl}
\affiliation{Department of Applied Physics, Eindhoven University of Technology, P.O. Box 513,
3500 MB Eindhoven, The Netherlands}
\author{Mihail I. Kotsev}
\affiliation{Department of Chemistry, Durham University, South Road, Durham DH1 3LE, United Kingdom}
\author{Mark A. Miller}
\email{m.a.miller@durham.ac.uk}
\affiliation{Department of Chemistry, Durham University, South Road, Durham DH1 3LE, United Kingdom}
\author{Paul van der Schoot}
\affiliation{Department of Applied Physics, Eindhoven University of Technology, P.O. Box 513,
3500 MB Eindhoven, The Netherlands}
\affiliation{Institute for Theoretical Physics, Utrecht University, Princetonplein 5, 3584 CC Utrecht, The Netherlands}

\begin{abstract}
	We investigate percolation in mixtures of nanorods in the presence of external fields that align or disalign the particles with the field axis.
	Such conditions are found in the formulation and processing of nanocomposites, where the field may be electric, magnetic, or due to elongational flow.
	Our focus is on the effect of length polydispersity, which -- in the absence of a field -- is known to produce a percolation threshold that scales with the inverse weight average of the particle length.
	Using a model of non-interacting spherocylinders in conjunction with connectedness percolation theory, we show that a quadrupolar field always increases the percolation threshold and that the universal scaling with the inverse weight average no longer holds if the field couples to the particle length.
	Instead, the percolation threshold becomes a function of higher moments of the length distribution, where the order of the relevant moments crucially depends on the strength and type of field applied.
	The theoretical predictions compare well with the results of our Monte Carlo simulations, which eliminate finite size effects by exploiting the fact that the universal scaling of the wrapping probability function holds even in anisotropic systems.
	Theory and simulation demonstrate that the percolation threshold of a polydisperse mixture can be lower than that of the individual components, confirming recent work based on a mapping onto a Bethe lattice as well as earlier computer simulations involving dipole fields.
Our work shows how the formulation of nanocomposites may be used to compensate for the adverse effects of aligning fields that are inevitable under practical manufacturing conditions.
\end{abstract}

\maketitle


	\section{Introduction}
		Composite nanomaterials have attracted a lot of attention due to their wide range of applications.
		The mechanical, thermal and electrical properties of a polymeric host medium can be greatly enhanced by the addition of a sufficient amount of nanofillers\,\cite{DresselhausCNTbook}.
		In order to design a material with the desired properties, it is crucial to understand and control the formation of a system-spanning network of nanofillers inside the host matrix, which happens above a critical concentration of filler material called the \textit{percolation threshold}\,\cite{Torquatobook}.
		For many technological applications, the percolation threshold is desired to be as low as possible, so as to not adversely affect other properties of the host material, such as mechanical stability, optical transparency and ease of processing.
		A case in point are carbon nanotubes, {which are} used in composites for electromagnetic interference shielding and transparent electrodes, {but} can strongly reduce the transparency of the material {if their concentration is too high}\,\cite{Kim2002, Shah2015}.
		
		Not surprisingly, a huge effort has been undertaken to remedy this problem and reduce the percolation threshold of carbon-nanotube-based materials.
		There are many factors that conspire against the formation of a conducting network, including the quality of the nanotubes, their mean aspect ratio, interactions and the processing of the composite\,\cite{KoningCNTbook, DeSimone1986_1, Vigolo2005, Deng2009, Kyrylyuk2008, Otten2012}.
		Processing steps like compression molding, spin coating, extrusion or drawing almost invariably cause some degree of alignment of the particles, and upon solidification the aligned structures are frozen in to remain present in the final product\,\cite{Grossiord2006}.
		Theory\,\cite{Balberg1984, Longone2012, Chatterjee2014, Otten2012}, experiments\,\cite{Du2005, Grossiord2006, Deng2009} and computer simulations\,\cite{Rahatekar2005, White2009, Kale2016} agree that particle alignment generally leads to higher percolation thresholds.
		A natural question that presents itself is: Can we mitigate the adverse effects of alignment on the percolation threshold by controlling the formulation of the composite?
		
		We argue that a sensible candidate for this would be the polydispersity level of the filler material.
		It is well known, theoretically and experimentally, that polydispersity has a significant impact on the percolation threshold\,\cite{Balberg1984, Balberg1986, Otten2009, Otten2011, Chatterjee2010, Nigro2013, Chatterjee2014, Meyer2015, Kale2015, Tkalya2014, Ambrosetti2010}. 
		Despite the strong influence of alignment and polydispersity, only limited effort has been untertaken to elucidate the interplay of these two effects.
		Dissipative Particle Dynamics simulations of aligned fillers suggest that less oriented short rods can facilitate network formation by connecting between more strongly aligned long ones\,\cite{Rahatekar2010}.
		This has been confirmed theoretically by mapping the continuum percolation problem onto percolation on a Bethe-lattice\,\cite{Chatterjee2014}.
			
		Applying continuum percolation theory, we show that the percolation threshold of polydisperse fillers can indeed be lower than that of the corresponding monodisperse ones.
		In fact, the known scaling of the percolation threshold with the inverse weight average of the particle length distribution is broken in the presence of an external field.
		Our conclusions confirm and generalise earlier findings and base them on a firm theoretical footing, which allows for natural extensions in other directions.
		One such extension of the theory would be to investigate the influence of confinement in thin films, which we intend to do in the near future.
		
		In Section \ref{sec:theory} we first outline the connectedness percolation formalism for length-polydisperse uniaxial particles with an arbitrary orientational distribution function. 
		Next, we rely on a Boltzmann weight to specify the orientational distribution in Section \ref{sec:field}, presuming a quadrupole alignment field that mimics the impact of electric, magnetic or elongational flow fields by explicitly coupling to the particle length.
		Details of our simulation method are given in Section \ref{sec:MC}, where we present a systematic and internally consistent way of determining the bulk percolation threshold in a finite size simulation. 
		This requires special attention if the system is anisotropic.
		Section \ref{sec:mono} focusses on the percolation threshold of monodisperse fillers, where we demonstrate that a quadrupole field always raises the percolation threshold and compare our theoretical predictions to Monte Carlo simulation results. 
		Our results are extended to polydisperse fillers in the Sections \ref{sec:bi} and \ref{sec:poly}.
		Using an example of a bidisperse system, we first show in Section \ref{sec:bi} that the universal scaling of the percolation threshold with the inverse weight average is broken in case the external field couples to the length of the filler particles.
		Next, we demonstrate how this can be exploited to lower the percolation threshold by changing the formulation of the composite.
		In Section \ref{sec:poly}, we give a generalised expression for the percolation threshold with an arbitrary length distribution of the fillers.
		Finally, in Section \ref{sec:dis}, we summarise our main results and set them side by side with previously published theoretical and simulation works.
		We further discuss the validity of our assumptions and provide suggestions for future directions.

		
	\section{Connectedness Percolation Theory \label{sec:theory}}		
		
		To investigate the percolation of rods in a composite material, we assume the network to be formed in the fluid stages of material processing.
		This way, the resulting particle configuration is in thermal equilibrium and can be studied theoretically using the framework of connectedness percolation theory, where connectedness is typically defined via a distance criterion.	
		Connectedness percolation theory is 
		based on the Ornstein-Zernike equation, which plays a central role in liquid state theory\,\cite{Bug1986, Coniglio1977, Hansen}.		
		It treats two-body connectedness probabilities analogously to liquid state correlation functions and has been widely applied in the context of spherical and non-spherical particles and mixtures thereof\,\cite{Torquatobook, Fantoni2007, Kyrylyuk2008, DeSimone1986_1, DeSimone1986_2}.
		In particular for slender rod-like particles, the agreement with results from Monte Carlo simulations is excellent\,\cite{Kyrylyuk2011, Schilling2015, Kale2015, Otten2009, Otten2011}.
		
		Here, we apply connectedness percolation theory to carbon nanotubes that we model as penetrable non-interacting spherocylinders of variable length and fixed width.
		In our model, two particles are directly connected if they overlap, which reduces the problem to purely geometric percolation.
		The rods present in our dispersion are modelled as perfect cylinders with a length $L_i$ and a constant diameter $\lambda$, end-capped with hemispheres of the same diameter.
		There is a fraction $x_i$ of rods of length $L_i$, where $i=1,2,...,n$.
		Here, $n$ denotes the number of components, which can, in principle, be infinite.
		We note that the diameter $\lambda$ of our model rods is not equal to the ``hard core'' diameter of the carbon nanotubes, but must somehow be related to the average tunneling distance of charge carriers in the host matrix.  
		We choose not to model these details explicitly and postpone a discussion of potential implications to Section \ref{sec:dis}.
		
		In order to determine the percolation threshold in the thermodynamic limit, we need to calculate the particle density at which the average cluster size, i.e., the mean number of connected particles, diverges.
		The mean cluster size
		\begin{align}
			S = \big\langle T_{i}(\bu) \big\rangle_{i, \bu} \label{S}
		\end{align}
		is equal to the average of all cluster sizes $T_{i}(\bu)$ of clusters that consist of at least one test rod of length $L_i$ and orientation $\bu$.
		Here, $\bu$ is the unit vector along the principal body axis, and $ \langle \dots \rangle_{i, \bu} = \int \dd \bu \sum_{i}  x_{i} \psi_{i}(\bu) (\dots)$ an average over the length distribution $x_i$ and orientational distribution $\psi_{i}(\bu)$.
		The orientational distribution function couples to the external field and may also be a function of the length of a particle.
		We discuss its functional form in the following Section.
		
		Within connectedness percolation theory, $T_{i}(\bu)$ needs to be solved from a generalised connectedness Ornstein-Zernike equation that takes the form\,\cite{Otten2009,Otten2011}
		\begin{align}
			  T_{i}(\bu)\!  =\! 1\! +\! \lim_{\bq\rightarrow 0}\rho \left\langle \hat{C}_{i j}^+(\bq,\bu, \bu') T_{j }(\bu') \right\rangle_{\! j,\bu'} \! . \label{polyOZE}
		\end{align}
		Here, $\rho$ denotes the overall number density of filler particles in the dispersion and $\hat{C}_{i j}^+ (
		,\bu,\bu')$ the spatial Fourier transform of the direct connectedness function $C^+$, with $\bq$ the wave vector.
		As, by definition, a rod is always connected to itself, the minimum cluster size is one, which leads to the first term on the right-hand side of eq.\,\eqref{polyOZE}.
		The second term denotes the number of other particles of length $L_j$ and orientation $\bu'$ that our test rod may additionally be connected to.
		In the limit of vanishing wave vector, we effectively average over all possible positions of other particles within the same cluster.
		
		To close this self-consistent integral equation, we invoke the second virial approximation, or chain-sum approximation, which becomes exact in the limit of infinite aspect ratios\,\cite{Schilling2015, Otten2011}.
		In practice, it turns out to be quantitative for aspect ratios $L / \lambda \gtrsim {400}$\,\cite{Schilling2015, Kale2016}, and is thus a suitable approximation for carbon nanotubes\,\cite{Ambrosetti2010}.
		Within the second virial approximation $\hat{C}_{i j}^+ (\bq,\bu, \bu') = \hat{f}_{i j}^+ (\bq,\bu, \bu')$\,\cite{Coniglio1977, DeSimone1986_1}, where $\hat{f}^+$ denotes the spatial Fourier transform of the connectedness Mayer function $f^+ = \exp(-\beta U^+)$ and $\beta$ the inverse thermal energy.
		The connectedness potential $U^+$ is zero if the particles are connected, and infinitely large otherwise.
		Thus, $\hat{f}_{i j}^+ (0,\bu, \bu')$ corresponds to the contact volume of two rods, i.e., the volume that the center of mass of one rod can trace out so that it still overlaps with a fixed second rod.
		
		The contact volume of two spherocylinders of lengths $L_i$ and $L_j$ and orientations $\bu$ and $\bu'$ is given by
		\begin{align}
		\begin{split}
			\hat{f}_{ij}^+(0,\bu,\bu') = \, & 2\lambda L_i L_j |\sin\gamma (\bu,\bu')|\\
				&+  \pi (L_i + L_j)\lambda^2 + \tfrac{4}{3} \pi \lambda^3 , \label{f}
		\end{split}
		\end{align}
		where $\gamma(\bu,\bu')$ is the angle between the main body axes of the rods\,\cite{Onsager1949, Balberg1984}.
		The impact of the $\lambda^2$ and $\lambda^3$ terms on the percolation threshold depends on the level of particle alignment as well as the aspect ratio of the rods.
		It turns out to be negligible for aspect ratios for which the second virial approximation is accurate\,\cite{Otten2011}, so we ignore these terms in most of our analyses.
		
		The mathematical problem that we seem to face is that we need to insert eq.\,\eqref{f} for the direct connectedness function in eq.\,\eqref{polyOZE} and solve for $T$.
		This is not quite true, since we are only interested in the average $S=\langle  T_{i}(\bu) \rangle_{i,\bu}$ of this function.
		If we average eq.\,\eqref{polyOZE}, we obtain $\langle  T_{i}(\bu) \rangle_{i,\bu} = 1 +\lim_{\bq\rightarrow 0}\rho \Big\langle\big\langle \hat{f}_{i j}^+(\bq,\bu, \bu') \big\rangle_{i,\bu} T_{j}(\bu')\Big\rangle_{j,\bu'}$.
		This equation still contains the function $T$ unless we take its average out of the convolution integral.
		Allowing for this \textit{ad hoc} approximation, we obtain the solution
			\begin{align}
				S = \left( 1-\rho\langle \langle \hat{f}^+ \rangle_{i,\bu}\rangle_{j,\bu'}\right) ^{-1},
			\end{align}
			which diverges at a critical particle density
			\begin{align}
				\rho_\pp =  \langle \langle \hat{f}^+ \rangle_{i,\bu}\rangle_{j,\bu'} ^{-1}. \label{PAA}
			\end{align}
		This result physically means that, on average, at least one rod per contact volume is required to form a system-spanning network.
		As the contact volume, and therefore the overlap probability of two rods, is maximal for isotropic configurations and minimal for perfectly parallel rods, any type of particle alignment is expected to increase the percolation threshold \cite{Balberg1984, Otten2012, Chatterjee2014}.
		This agrees with what is seen in experiments\,\cite{Du2005} and in computer simulations\,\cite{Balberg1984_2, White2009, Kale2016}.
		
		However, this pre-averaging approximation\,\eqref{PAA} turns out to be exact only for isotropic and perfectly parallel monodisperse rods\,\cite{Otten2011}.
		For polydisperse filler particles, it fails to introduce the correct moments of the particle length distribution even in isotropic configurations\,\cite{Balberg1984, Balberg1986, Otten2009, Otten2011, Chatterjee2010, Chatterjee2014}, where the correct percolation threshold reads $\rho_\pp = 2/  \pi \lambda \langle L^2 \rangle $\,\cite{Otten2011}.
		Below, we will make a detailed comparison between the pre-averaging approximation and a systematic approximation scheme.
		But first, let us discuss in what way the orientational distribution function of the rods depends on the strength of an external alignment field, which we choose to be of the quadrupole type.
		
	\section{The external alignment field \label{sec:field}}
		In the case of non-interacting rods in an external field, the orientational distribution function $\psi$ needed to evaluate the averages $\langle \dots \rangle_{\bu}$ is given by the normalised Boltzmann factor of the external potential $U$:
		\begin{align}
			\psi = N^{-1} \exp\left( -\beta U \right).
			\label{Boltzmann}
		\end{align}
		We presume the aligning field $U$ to be of the quadrupole type, so that 
		\begin{gather}
			\beta U = K \cos^2\theta \label{energy},
			\shortintertext{and the normalisation factor becomes}
			N=2 \pi \sgn(K)\sqrt{\pi/K}\erf(\sqrt{K}),
		\end{gather}
		where $\theta$ is the polar angle of a rod with the field direction, and $K$ denotes a dimensionless measure {of} the field strength.
		In the case of negative field strengths, rods are {preferentially} aligned parallel to the field direction, whereas for positive values they orient perpendicular to that.
		The degree of alignment can be quantified by the nematic order parameter $\langle P_2 \rangle = \left( 3 \langle \cos^2 \theta \rangle - 1\right)/2 $, which we
		can calculate exactly from the normalisation as $\langle P_2(K) \rangle = - \tfrac{3}{2N} \dd N / \dd K - \tfrac{1}{2}$.
		It is $1$ for perfectly parallel rods, which corresponds to the field strength $K\rightarrow  -\infty$, and zero for isotropic rods, i.e., for vanishing fields $K=0$.
		If $K\rightarrow \infty$, the rods are perfectly ``disaligned'', i.e., they are isotropically oriented perpendicular to the field direction with $\langle P_2 \rangle = -1/2$.\\
		
		Experimentally, the alignment of carbon nanotubes can be caused by electric\,\cite{Bubke1997, Yamamoto1998, Chen2001, Martin2005, Brown2007, Zhu2009} or magnetic fields\,\cite{Fujiwara2001, Zaric2004}, elongational flow\,\cite{Wang2008, Du2005, Aguilar2016}, shear\,\cite{Wescott2007} or liquid crystalline solvents\, \cite{Lagerwall2008review, Lagerwall2016book, PopaNita2008}.
		The exact realisation of the field strength parameter $K$ depends crucially on the type of field applied.
		A rod submerged in a thermotropic nematic host medium, for instance,  feels a quadrupole type potential,  with $K=L\lambda W \pi / 3$ under weak anchoring conditions, where $W$ is the average anchoring energy\,\cite{PopaNita2008}. 
		In the case of an electric field, the dimensionless field strength is given by $K = - \beta \Delta \alpha E^2 / 2 $, where $E$ is the electric field strength and $\Delta \alpha$ the rod's polarisability anisotropy. 
		Similarly, the magnetic quadrupole field gives $K = - \beta \Delta \chi H^2 / 2$ with the magnetic field strength $H$ and the susceptibility anisotropy $\Delta \chi$.
		Using these expressions, we assume that the rods do not possess a permanent dipole moment and neglect dipole-dipole interactions, which are known to cause lateral clustering or chain formation\,\cite{Martin2005, Alvarez2012, Alvarez2013}.
		
		An effective quadrupole field can, in principle, also be realised by an elongational flow field\,\cite{Khokhlov1982, DoiEdwards}, which is the only flow field that allows for a quasi-static treatment.
		In the case of a uniaxial elongation or compression, the field strength is defined as $K = -3\dot{\epsilon} / 4 D_\mathrm{r}$, where $ D_\mathrm{r}$ denotes the rotational diffusion coefficient of a straight rod and $\dot{\epsilon}$ is the strain rate, which is positive for elongational flow and negative for compression \cite{DoiEdwards, Aguilar2016}.
		
		It is crucial to note that the dimensionless field strength $K$ {typically} increases with the rod length, and that, as a result, alignment fields affect long particles more strongly than short ones.
		How strongly $K$ changes with $L$ depends on the type of field applied.
		The surface energy of a rod submerged in a thermotropic nematic solvent, for instance, increases linearly with the surface area and thus with the length of the rod\,\cite{Brochard1970, PopaNita2008}.
		Also in the case of an electric or magnetic field, we presume the rod's polarisability and susceptibility anisotropy to be proportional to the polarisable volume and therefore to the rod length $L$, at least in the limit of large enough aspect ratios.
		In practice, this turns out to be accurate both for metallic and semiconducting carbon nanotubes in magnetic\,\cite{Zaric2004} and electric fields\,\cite{Benedict1995, Venermo2005}, again provided that the particles are long enough.

		The rotational diffusion coefficient $D_\mathrm{r}$, which becomes relevant in the case of a hydrodynamic flow field, depends on the particle length in a non-trivial way.
		It is well known that, in dilute dispersions, $D_\mathrm{r} \propto L^{-3}$ if we neglect the logarithmic correction\,\cite{DoiEdwards}. 
		In the semidilute regime, where the rods are entangled, it has been proposed that $D_\mathrm{r}\propto L^{-9}$\,\cite{DoiEdwards, Keep1985}, even though the exact scaling has been subject to extensive debate.
		Other theoretical models\,\cite{Chung1989} and experimental work\,\cite{Maguire1980} find the relations $D_\mathrm{r}\propto L^{-7}$ and  $D_\mathrm{r}\propto L^{-5.7}$ respectively.
		While the exact scaling of $D_\mathrm{r}$ in the semidilute regime remains a matter of contention, it is clear that the rotational diffusion coefficient decreases with some power of the length, and that this power is larger than 3.
		
		We conclude that the orientational distribution function depends on the particle length $L_i$ to a certain power $P$ that is characteristic for the type of field applied.
		{While this power is positive in most physical situations, it is in principle also possible to fabricate nanoparticles with $P<0$, where short rods align more strongly than long ones\,\cite{vanRhee2013}. 
		In order to account for the coupling between the particle length and the external field, we} write the dimensionless field strength as
		\begin{align}
			K = K_0 \left( \frac{L_i}{\lambda} \right) ^P \label{K}
		\end{align}
		with $K_0$ a bare field strength parameter that we do not specify further, and $L_i/\lambda$ the aspect ratio of the rod.
		For weak fields, $\langle P_2 \rangle_i \sim -\tfrac{2}{15} K_0 \left( \tfrac{L_i}{\lambda} \right)^P + \cdots$, showing that the larger $P$ the weaker the field needs to be in order to obtain a significant degree of order.
		As we shall see,  this will have a significant impact on the percolation threshold for bidisperse mixtures of short and long rods, both in theory and simulations.
		However, before proceeding to the predictions of our theory, we first discuss our Monte Carlo algorithm for anisotropic rod mixtures in the following Section.


	\section{Monte Carlo Simulations\label{sec:MC}} 
To simulate the model of fully penetrable, non-interacting spherocylinders in a quadrupole field as introduced in Section \ref{sec:field}, we generate explicit Boltzmann-distributed configurations of the rods and perform a cluster analysis to detect percolation.
		Since the positions and orientations of ideal objects are completely uncorrelated, independent configurations of spherocylinders are readily generated by random sequential insertion into the simulation cell {without the need for cluster moves or neighbor lists}.  
		The centers of the spherocylinders are uniformly randomly distributed within a cubic periodic cell of length $L_{\rm box}$.  
		The external field is oriented in the $z$ direction, and the azimuthal angles of the spherocylinders about this axis are uniformly randomly distributed. 
		The thermal distribution of polar angles defined by eqs.~\eqref{Boltzmann} and \eqref{energy} can, in principle, be generated by transformation of a uniform random deviate.  
		However, to do so for both aligning and disaligning fields requires the efficient and accurate evaluation of the inverse error function for real and imaginary arguments, respectively.  
		Here, we take the simpler approach of sampling the polar angle of each spherocylinder by Metropolis Monte Carlo steps in $\cos\theta$.  
		For a smooth one-dimensional function like eq.\,\eqref{energy} such steps rapidly randomize the orientations within the required thermal distribution.

		Let $p(\phi,L;L_{\rm box})$ be the probability that a configuration in a simulation of spherocylinders of length $L$ at packing fraction $\phi=N\pi\lambda^2(2\lambda+3L)/(12L_{\rm box}^3)$ contains a percolating cluster, where $N$ is the number of spherocylinders.  
		In the thermodynamic limit $L_{\rm box}\to\infty$, this probability would be a sharp step function, $p(\phi,L;\infty) = \Theta\left(\phi-\phi_{\rm p}(L)\right)$ but for finite $L_{\rm box}$, the probability becomes a smooth sigmoidal function of $\phi$.  
		Skvor {\em et al.} \cite{Skvor07a} have shown that $\phi_{\rm p}$ can nevertheless be extracted from simulations by exploiting the scaling properties of $p(\phi,L;L_{\rm box})$ with $L_{\rm box}$.  
		To do this, it is essential to define a percolating cluster by a ``wrapping'' criterion, which requires that any particle in the cluster is connected to its periodic images by a contiguous path of contacts through the cluster; 
		it is not sufficient for the cluster merely to have a physical extent greater than $L_{\rm box}$.  
		With this definition, curves of $p(\phi,L;L_{\rm box})$ as a function of $\phi$ have a common crossing point for all values of $L_{\rm box}$.  
		Since this property must also apply in the thermodynamic limit, the crossing point derived from two different values of $L_{\rm box}$ gives an accurate estimate of $\phi_{\rm p}$.

		For systems of spherical \cite{Miller09c} or isotropically oriented particles \cite{Schilling07a}, the common crossing point tends to lie just below a probability of $0.5$.  
		Hence, in such cases, a reasonable esimate of $\phi_{\rm p}$ can be obtained from simulations at a single value of $L_{\rm box}$ by locating the point where the probability passes through $0.5$.  
		However, in the present study, the (dis)alignment of particles by the external field leads to a significant change in the sigmoidal shape of $p(\phi,L;L_{\rm box})$.  
		Curves from different $L_{\rm box}$ still retain a common crossing point, but the value of the probability at the crossing shifts further from $0.5$ as the magnitude of the field is increased.  
		Hence, in the anisotropic system, it is essential to determine $\phi_{\rm p}$ from a scaling analysis and not to rely on an arbitrary threshold in the probability.
Failing to apply the scaling analysis can even lead to the qualitatively incorrect prediction that weak alignment of the rods slightly lowers the bulk percolation threshold.

		\begin{figure}
			\includegraphics[width=1.02\linewidth]{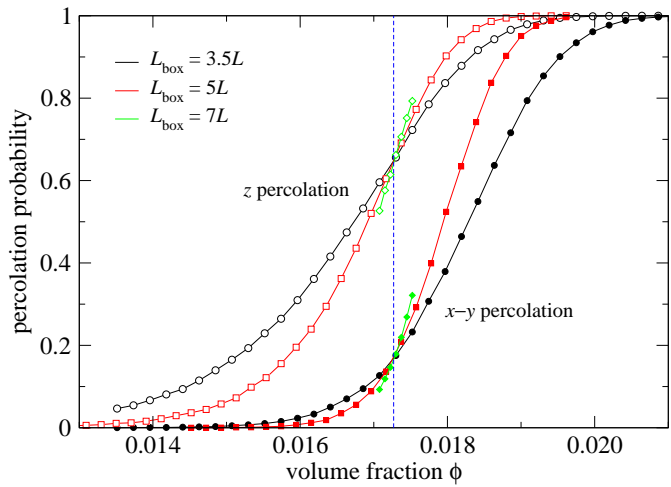}
			\caption{Percolation probability of ideal spherocylinders of aspect ratio $L/\lambda=50$ in an aligning field $K=-5$ from Monte Carlo simulations, {sampling 50,000 configurations per point}.  
				Open symbols show the probability of observing a cluster that wraps the cubic simulation cell in the direction $z$ of the field, while closed symbols refer to wrapping in at least one of the perpendicular directions.  
				For each direction, probabilities for three cell lengths $L_{\rm box}$ are shown.  
				Each set of curves has a common crossing point and the two crossing points occur at the same packing fraction (dashed vertical line), defining the percolation threshold in the bulk limit.}
			\label{fig:cumulants}
		\end{figure}

		Figure \ref{fig:cumulants} illustrates the use of the scaling analysis to determine the percolation threshold of spherocylinders in an aligning field.  
		Increasing $L_{\rm box}$ results in sharper sigmoidal curves of $p(\phi,L;L_{\rm box})$ and reveals a common crossing point.  
		In the figure, two percolation probabilities are shown: one for wrapping across the periodic boundary conditions in the direction of the field $z$ and the other for wrapping in at least one of the orthogonal directions ($x$--$y$).  
		{Overall,} the two sets of curves are mutually displaced with respect to packing fraction.
		This shows that, {for a finite, cubic sample} at a given packing fraction, the probability of percolation in the direction of the field is greater than in the orthogonal directions.  
		This observation gives the impression that the material has an anisotropic percolation threshold.  
		However, {in addition to the overall displacement between the two sets of curves, there is a difference in their sigmoidal shapes, and this causes the crossing points to occur at different percolation probabilities.  This change in shape is visible in Fig.~\ref{fig:cumulants}, where the curves for percolation in the $z$ direction initially rise less steeply than those for $x$--$y$ percolation.}  The scaling analysis reveals that, {as the field strength increases, the crossing points for parallel and perpendicular percolation shift in opposite directions on the probability axis.  This effect exactly counteracts the overall horizontal displacement of the curves and results in the crossing points occurring at the same value of the packing fraction}.
		Hence, although the distribution of particle orientations is anisotropic, the percolation threshold itself is isotropic in the bulk limit.  

		The origin of the apparent discrepancy in finite-sized simulations is that the correlation lengths are different in the directions parallel and perpendicular to the field.
		As a result, clusters below the percolation threshold are non-spherical on average.  
		Although the correlation lengths grow at different rates as the percolation threshold is approached from below, it has been predicted theoretically that they should diverge at the same packing fraction \cite{Otten2012}.  
		It has also been shown in simulations of hard ellipsoids that the difference between the parallel and perpendicular percolation thresholds decreases systematically as the simulation size is increased \cite{Kale2016}.  
		We believe that our simulations are the first to show explicitly that the difference vanishes completely in the bulk limit, giving an isotropic percolation threshold.
		
		The same analysis may be applied in the presence of polydispersity in addition to the external field.  
		In the simulations of bidisperse mixtures, each rod is randomly assigned one of two lengths according to the desired target distribution in each generated configuration.  
		Hence, the proportion of each species fluctuates slightly from one configuration to another, but the average over many configurations is equal to the target average.  
		In the present work, all points in $p(\phi,L;L_{\rm box})$ are obtained from at least {5,000} independent configurations.
		To obtain $\phi_{\rm p}$ from systems of two different sizes, we locate the crossing point by linear interpolation between points typically differing by about 0.1 in probability.  The statistical uncertainty in Monte Carlo data presented in the figures is below $0.2\%$.

		At low field strengths, where the distribution of polar angles is broad, cubic boxes were used with $L_{\rm box}$ equal to $3.5$ and 5 times the length of the longer species in the mixture.
		At higher field strengths, where at least the longer species is strongly aligned along the $z$ axis, elongated cells of dimensions $2L\times2L\times4L$ and $3L\times3L\times6L$ were used.  
		{Note that these two elongated cells have identical aspect ratios, which is essential in order to obtain a common crossing point for systems of different size, analogous to Fig.~\ref{fig:cumulants}.  
		For a family of cells with the same aspect ratio but different sizes, the curves of $p(\phi)$ differ only in their width, and can be scaled about their crossing point to collapse onto each other \cite{Skvor07a}.
		Moving from a cubic to an elongated simulation cell alters the shape of $p(\phi)$, creating a new set of scalable curves when the size of the cell is changed at fixed aspect ratio.  These curves have a common crossing point at a different value of the probability compared to the cubic cells.  However, the packing fraction at which the crossing occurs is the same for a family of cubic cells as for a family of elongated cells.}
		Hence, an elongated cell reaches the same result as a cubic cell, but is more efficient for highly aligned systems {because it contains fewer particles}.
\par
		Although it is computationally trivial to generate configurations of ideal rods of any length, it does become more demanding to evaluate $\phi_{\rm p}$ for longer rods using the methods described above. 
		 This is because more rods must be simulated when $L$ is greater.  In a cubic box with edge measured as a multiple of $L$, the number of particles for a given number density increases as $L^3$.  
		 Counterbalancing this increase is the approximate scaling of $\phi_{\rm p}$ with $1/L$ and the fact that the volume of an individual spherocylinder is approximately proportional to $L$.  
		 Hence, the number of rods required to follow the percolation threshold increases roughly linearly with $L$ overall.  
		 However, the time taken for the cluster analysis scales approximately as the square of the number of particles present.  
		 Therefore, the computational cost increases roughly as $L^2$.
		 {The largest systems simulated in the present work involve more than 60,000 rods.  
		 The cluster analysis of a single configuration may therefore include up to $1.8\times 10^9$ checks for pairwise overlaps between rods.  
		 Hence, these simulations are unusual in that it is far more costly to analyze the configurations than to obtain an ergodic sample of configurations in the first place.}

	\section{Monodisperse Rods\label{sec:mono}}
		
		Before discussing how polydispersity and an external field affect the percolation threshold of ideal spherocylinders, we first focus on the external field alone.
		This allows us to explain more clearly the way we calculate the percolation threshold from the governing equations and to discuss the strengths and deficiencies of our theoretical approach.
		
		From equations\,\eqref{S} and \eqref{polyOZE}, we calculate the percolation threshold $\rho_\mathrm{p}$ of monodisperse fillers in an external field by making use of the cylindrical symmetry of the problem.
		Because the orientational distribution function $\psi(\theta)$ is independent of the azimuthal angle $\varphi$ of the rods, the $\varphi$-average only requires the integration of $\left| \sin \gamma(\bu,\bu') \right|$.
		Expanding the integral kernel in Legendre polynomials $P_{2n}$ and using the addition theorem for spherical harmonics\,\cite{Jackson, Odijk1988}, we obtain
		\begin{align}
			\hspace{-4pt}\int_0^{2\pi}\hspace{-8pt}\left| \sin \gamma{\small (\bu,\bu')} \right| \! \dd\varphi = 2\pi \hspace{-2pt} \sum_{n=0}^{n_\text{max}} \hspace{-2pt}  d_{2n} P_{2n}(\cos\theta) P_{2n}(\cos(\theta'), \label{additiontheorem}
		\end{align}
		with the coefficients $d_0 = \pi \big/ 4$ and, for $n > 0$,
		\begin{align} 
			d_{2n}=- \pi (4n+1) \frac{(2n-3)!!(2n-1)!!}{2^{2n+2}n!(n+1)!},
		\end{align}
		where $!!$ denotes the double factorial\,\cite{Jackson}.
		In principle, the upper bound $n_\text{max} \rightarrow \infty$, but in practice we truncate the series and choose $n_\text{max}$ to produce a sufficiently accurate prediction for the percolation threshold.
		Inserting this into eq.\,\eqref{polyOZE} results in an expression for $T$ which only depends on the polar angle $\theta$,
		\begin{align}
			T( \theta) = 1 + 2 \rho \lambda  L^2 \sum_{n=0}^{n_\mathrm{max}}d_{2n} P_{2n} (\cos\theta) \left\langle P_{2n} (\cos\theta')  T(\theta')\right\rangle_{\theta'} \label{T} .
		\end{align}
		Here, we {have} dropped the indices $i$ and $j$ to stress the monodisperse nature of the filler particles and {have} ignored the end-cap corrections to the contact volume.
		
		In order to solve this self-consistent equation and determine the percolation threshold for arbitrary field strengths, we need an expression for the moments $\left\langle P_{2n} (\cos\theta')  T(\theta')\right\rangle_{\theta'}$.
		Multiplying eq.\,\eqref{T} with an even Legendre polynomial $P_{2l}(\cos\theta)$ and subsequent averaging over $\theta$ leads to the following set of linear equations with a number $n_\text{max}+1$ of unknowns:

		\begin{align}
			\langle &P_{2l}(\cos\theta) T(\theta) \rangle_{\theta} = \left\langle P_{2l}(\cos\theta) \right\rangle_{\theta} + 2 \rho L^2 \lambda \\
			&\times \! \sum_{n=0}^{n_\text{max}}  d_{2n} \left\langle P_{2l}(\cos\theta) P_{2n}(\cos\theta) \right\rangle_{\theta} \left\langle P_{2n}(\cos\theta') T(\theta')\right\rangle_{\theta'}.\nonumber
		\end{align}
		
		Formally, the solution of this set of equations can be obtained by inverting the $(n_\text{max} +1)^2$-matrix. 
		To determine the critical density $\rho_\pp$ at which the cluster size $S$ diverges, however, an exact expression for $S$ is not needed.
		Instead, it is sufficient to set the determinant of the matrix equal to zero, which we do analytically using Wolfram Mathematica\,\cite{mathematica}.
		The resulting general expression for the percolation threshold is unwieldy so we do not reproduce it here.
		Instead, we illustrate our findings graphically.
		
		\begin{figure}
			\includegraphics[width=1.02\linewidth]{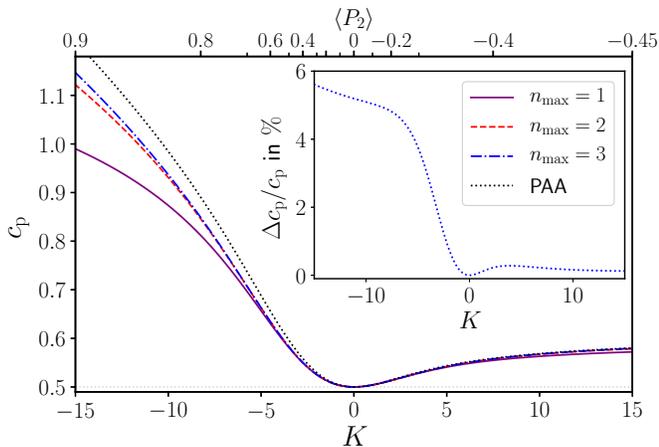}
			\caption{
				Percolation threshold of a monodisperse dispersion of ideal rods in terms of the dimensionless concentration {$c = \pi L^2 \lambda \rho / 4$} versus the field strength $K$ in the limit of infinite aspect ratio $L/\lambda \rightarrow \infty$.
				Here, $L$ denotes the length of the rods, $\lambda$ their width, and $\rho$ the number density.
				Solid, dashed and dash-dotted lines: solutions of eq.\,\eqref{T} with $n_\text{max}=\{1,2,3\}$. 
				Dotted line: pre-averaging approximation, eq.\,\eqref{PAA}.
				The non-linear top axis shows the nematic order parameter $\langle P_2\rangle$ of the rods that depends on the external field applied.
				Inset: relative error of the pre-averaging approach (PAA) with respect to the $n_\text{max}=3$ full solution.
				}
			\label{fig:orders}
		\end{figure}
		
		Introducing the dimensionless filler concentration {$c = \phi  L  / \lambda = \rho \lambda L^2 \pi / 4 $}, we scale out the known universal length dependence of the zero-field percolation threshold $\phi_\pp = \lambda / 2 L$.		
		In Figure \ref{fig:orders} we compare the percolation threshold within the pre-averaging approximation, eq.\,\eqref{PAA}, to solutions obtained with the full Ornstein-Zernike equation \eqref{T} for orders up to $n_\text{max}=3$.
		Note that this figure is universal, for the aspect ratio of particles is implicit in the scaled concentration and the field strength.
		Indicated also is the order parameter shown on the upper horizontal axis.
		The figure shows that $n_\text{max}=1$ is accurate within one per cent for order parameters $-0.44 \leq \langle P_2 \rangle \leq 0.66$.		
		The relative error of the percolation threshold obtained for a truncation after $n_\text{max}=2$ with respect to the solution for $n_\text{max}=3$ is found to be lower than $1\%$ for nematic order parameters $-0.5 \leq \langle P_2 \rangle \leq 0.86$ and stays below $2.2\%$ up to $\langle P_2 \rangle = 0.9$.
		This covers the range of experimentally accessible order parameters for rod dispersions\,\cite{Nastishin2005, Horowitz2005, Ould2013, Scalia2008, Scalia2010, Dierking2005, Lynch2002}.
		For this reason, and for computational simplicity, we ignore contributions of order $n_\text{max}=3$ and higher.
		
		Figure \ref{fig:orders} demonstrates that the pre-averaging approximation consistently overestimates the percolation threshold and is off by up to $6\%$ when compared to the full Ornstein-Zernike solution with $n_\mathrm{max} = 3$, for the range of field strengths probed.
		However, the pre-averaging approximation, whilst not so accurate for aligning fields, is remarkably accurate for disaligning fields.
		This is, in fact, not entirely surprising given that it is exact for fully isotropic configurations, and that in the disaligned state most of the rods are isotropically oriented perpendicular to the field direction.
		
		The exact prediction for isotropic configurations ($K=0$) is $c_\pp = 1/2$ if we ignore the end-cap contributions to the contact volume.
		For perfectly disaligned rods ($K\rightarrow \infty$) we find $c_\pp = \pi^2/16$, again up to leading order in the aspect ratio.
		In the limit ${K\rightarrow - \infty}$, however, we can no longer ignore the contributions of the end caps giving rise to a percolation threshold $c_\pp = (8 \lambda / L  +  4 \lambda^2 / 3 L ^2)^{-1} \sim L/8\lambda$ to leading order in the aspect ratio. 
		This implies that the packing fraction $\phi_\pp = c_\pp \lambda/L \sim 1/8$ at percolation for infinite aligning fields becomes an invariant of the aspect ratio of the particles.
		Of course, the second virial approximation breaks down in this limit. 
		In spite of this, we do expect that the packing fraction at percolation should remain an invariant of the aspect ratio also within more sophisticated closures.
		
		\begin{figure}
			\includegraphics[width=1.02\linewidth]{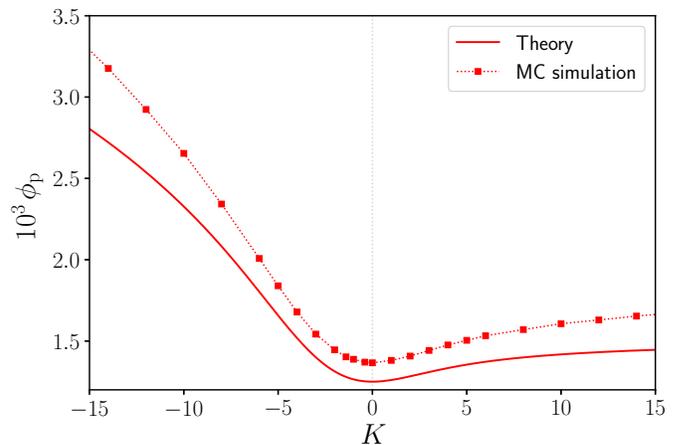}
			\caption{Direct comparison between the percolation threshold obtained with connectedness percolation theory (solid line) and Monte Carlo simulation results (full squares with dashed line) for rods of aspect ratio $L/\lambda = {400}$. The apparent deviations originate from the use of the second virial approximation in the theory, as described in the main text.}
			\label{fig:monoTheoSim}
		\end{figure}
		
		As already mentioned, the second virial approximation loses accuracy for aspect ratios below $\sim {400}$\,\cite{Schilling2015}.
		This is exemplified in Figure \ref{fig:monoTheoSim}, where we directly compare the theoretical prediction of the percolation threshold in terms of the filler fraction $\phi_\pp = \rho_\pp\cdot \pi\lambda^2(2\lambda+3L)/12$ to our Monte Carlo simulation results for the aspect ratio $L/\lambda = {400}$.
		Even though the qualitative agreement is very good, the simulation curve is shifted upwards with respect to the theoretical prediction, and the disagreement increases with increasing field strengths.
		This is not entirely surprising, because the contribution of higher virials becomes more significant the shorter and the more aligned the rods are\,\cite{Straley1973, Mulder1985, Mulder1987, Frenkel1987}.
		A calculation on the third virial level\,\cite{Otten2011, Straley1973}, which we do not discuss here in detail, raises the theoretical percolation threshold in isotropic dispersions by about {$1.2\,\%$} , leading to a better agreement between theory and simulations.
		
		In passing, we note that while our theory is the most accurate in the slender rod limit, this limit proves difficult to achieve in simulations, even for ideal particles, as discussed in Section \ref{sec:MC}.
		For this reason, we compare our predictions for bidisperse rods with simulations only for relatively modest aspect ratios in the following Section.

	\section{Bidisperse Rods\label{sec:bi}}
		
		To study the combined effect of length polydispersity and an external field, in this Section we specifically consider an example of bidisperse filler particles with fixed lengths $L_1$ and $L_2$ and number fractions $x_1$ and $x_2 = 1-x_1$ respectively.
		We focus on a bidisperse mixture due to its most noticeable effect on the percolation threshold compared to a continuous distribution with the same mean length, as shown in\,\cite{Otten2009, Otten2011}.
		The results for the more general case of an arbitrary length distribution are discussed in Section \ref{sec:poly}.
		
		Employing eq.\,\eqref{polyOZE} and the second virial approximation, the average cluster size in a polydisperse system reads
		\begin{align}
			\begin{split} 
			T_i(\theta)= & 2 \rho \lambda \sum\nolimits_n d_{2n}  L_i  P_{2n}(\cos\theta) \\
			& \times \left\langle L_j P_{2n}(\cos\theta') T_{j(\theta')}\right\rangle_{j \theta'} + 1,
			\end{split}
		\end{align}
		where we have again neglected the end caps and applied the addition theorem for spherical harmonics, eq.~\eqref{additiontheorem}.
		The moments of the function $T$ are then defined by the set of linear equations
		\begin{align}
			\langle L_i P_{2l}&(\cos\theta) T_{i\gamma}(\theta) \rangle_{i\theta} \nonumber\\
			=&\, 2 \rho  \lambda \sum\nolimits_n  d_{2n} \left\langle L_i^2 P_{2l}(\cos\theta) P_{2n}(\cos\theta) \right\rangle_{i\theta} \nonumber \\
			& \times \left\langle L_j P_{2n}(\cos\theta') T_{j}(\theta')\right\rangle_{j \theta'} \label{polyMatrix} \\	
			&+\left\langle L_i P_{2l}(\cos\theta) \right\rangle_{i\theta} \nonumber.
		\end{align}
		For the bidisperse case, $i,j \in \{1,2\}$, while for more general compositions $i,j \in \{1,2, \dots \}$.
		We determine the percolation threshold of polydisperse fillers following a similar procedure as described in the previous Section.		
		
\begin{figure}
			\includegraphics[width = \linewidth]{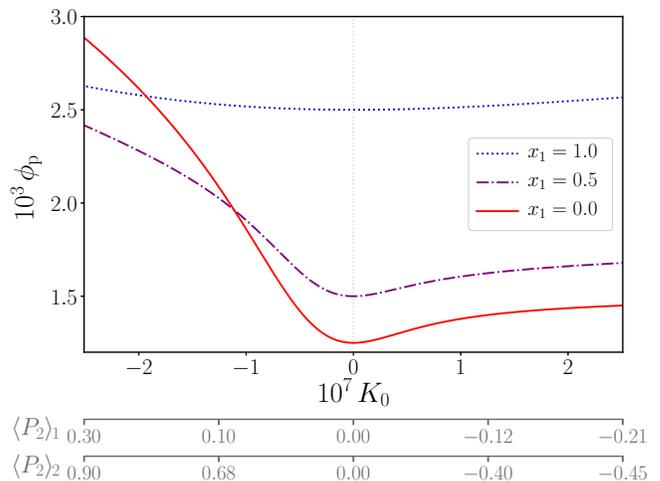}
			\caption{Critical volume fraction $\phi_\mathrm{p}$ versus the external field strength $K_0$ in a field with cubic length coupling, $P=3$ for the number fractions $x_1 = \{0, 0.5, 1\}$ and $x_2 = 1-x_1$ respectively.
				The aspect ratios are $L_1 / \lambda = {200}$ and $L_2 / \lambda = {400}$.
				Indicated are also the nematic order parameters $\langle P_2 \rangle$ for the two lengths of rod.
				Curves with different formulations cross each other.
				In the regime of strong aligning fields, the bidisperse system exhibits the lowest percolation threshold, as short rods can act as more isotropic linkers between the strongly aligned long ones.
				} 
			\label{fig:biP3}
		\end{figure}

		The trivial case is that where the field coupling parameter $P$ vanishes, i.e., when the field is independent of the particle length.
		In that case, the effect of polydispersity is entirely scaled out by the definition of the dimensionless concentration {$c = \pi \rho \lambda \langle L_i^2\rangle_i / 4$}, and we recover the universal curve for the percolation threshold that coincides with the solution for the monodisperse rods shown in Figure \ref{fig:orders}.
		For {$P\neq 0$}, however, this universality is broken by the length-field coupling, implying that the known scaling of the percolation treshold with the weight average of the length distribution does not hold anymore.
		This has the important consequence that the percolation threshold of a mixture can be lower than that of the individual constituents.
		We illustrate this in Figure \ref{fig:biP3} for the case $P=3$, mimicking the impact of an elongational flow field.
		Here, as in all figures to follow, we set $n_\text{max}=2$ in equation \eqref{additiontheorem}.
		As already alluded to, we expect this to be accurate to within a per cent for order parameters in the range $-0.5 \leq \langle P_2 \rangle \leq 0.86$.
		
		In Figure \ref{fig:biP3} we show the volume fraction $\phi_\pp$ at percolation as a function of the bare field strength $K_0$, as defined in eq.~\eqref{K}, for mixtures of rods of aspect ratios $L_1 / \lambda = {200}$ and $L_2 / \lambda = {400}$.
		Indicated are also the nematic order parameters $\langle P_2 \rangle$ of the two types of particle for our choice of $P=3$, confirming that longer particles are more susceptible to the effect of the orienting field than the short ones.
		We observe the following: 
		
		\begin{enumerate}
			\item For zero field, the dispersion containing \textit{only} the short rods exhibits the highest percolation threshold, whereas the one containing \textit{only} long rods percolates at much lower volume fractions, as expected;
			\item This remains true for disaligning fields, no matter how strong they are, and for aligning fields, provided that they are not too strong;
			\item For sufficiently strong aligning fields, however, we find the opposite: shorter rods form a percolating cluster at lower concentrations, on account of them not being as strongly oriented;
			\item The percolation threshold of a 50-50 mixture lies between that of the pure components unless the alignment field is sufficiently strong. In that case, the mixture has the lowest percolation threshold.
		\end{enumerate}
		
		Two comments are in order at this point.
		First, the monodisperse cases $x_1=0$ and $x_1=1$ can be described universally, even for $P>0$, by defining the appropriate concentration scale $c_\pp$ and the field strength
		\begin{align}
			\tilde{K}=K_0 \langle L_i^P \rangle /\lambda^P , \label{Kwiggle}
		\end{align}
		which we choose this way in order to compare mixtures with the same interaction energy per particle.
		The existence of a universal curve describing the connection between $c_\pp$ and $\tilde{K}$ does not apply for any other values of $x_1$.
		
		Second, that a mixture of long and short rods may have a lower percolation threshold than the individual species is due to a cooperative (synergetic) effect\,\cite{Kyrylyuk2011}.
		In this case, it is caused by short rods acting as {more isotropic} connectors between the strongly aligned long ones.
		The long rods contribute to the network by making long-range connections between separated clusters of short rods that otherwise would not form a system-spanning network.
		The importance of this last point has perhaps not been fully appreciated\,\cite{Rahatekar2010, Chatterjee2014}.
		
		As we shall see below, for this mechanism to work the rods need to have a sufficient difference in order parameter at a given field strength.
		Whether or not the mixture has the lowest percolation threshold depends on the coupling strength $P$, the number fraction $x_1$ and the length ratio $L_2 / L_1$.
		{It turns out that the cooperative network formation described above can only occur if long rods are more strongly aligned than short ones, i.e., for field-coupling parameters $P>0$. 
		In the case of $P<0$, turning on an external alignment field amplifies the advantage of the longer species in forming a network, which is already present in isotropic configurations.
		As a result, for negative field-coupling parameters $P$, the mixture containing only the longest species always exhibits the lowest percolation threshold.}
		
		Before {analysing the conditions for cooperative network formation in} more detail, we first discuss the comparison of our connectedness percolation theory with results from Monte Carlo simulations.
		
		\begin{figure}
			\includegraphics[width=0.97\linewidth]{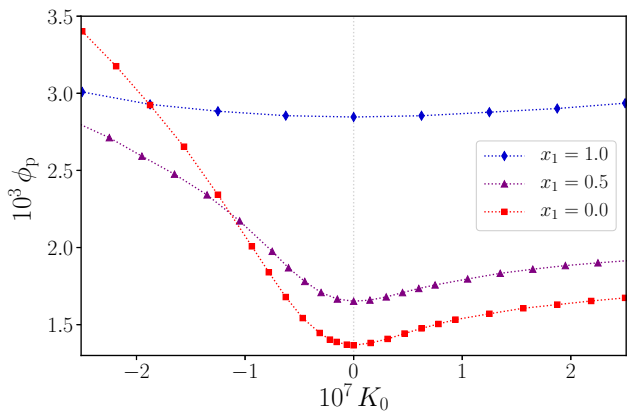}
			\caption{Percolation threshold in terms of the filler fraction obtained by Monte Carlo simulations for a field with cubic length coupling, $P=3$. 
				The aspect ratios are $L_1 / \lambda = {200}$ and $L_2 / \lambda = {400}$, with the number fractions $x_1 = \{0, 0.5, 1\}$ and $x_2 = 1-x_1$ respectively.}
			\label{fig:simCrossing}
		\end{figure}
		
		In Figure \ref{fig:simCrossing} we show our Monte Carlo simulation results for the percolation threshold of the same binary mixture of rods and the same field coupling parameter $P=3$.
		As already discussed in Section \ref{sec:mono}, the absolute values for the percolation threshold in theory and simulations are shifted with respect to each other.
		Despite this, we again find the cooperative effect in our Monte Carlo simulations, in very good qualitative agreement with the theory. 
		In fact, the field strengths below which the 50-50 mixture has a lower percolation threshold than the pure rods even agree well quantitatively;
		they {differ by less than 3}$\%$.
		
		\begin{figure}
			\includegraphics[width=0.97\linewidth]{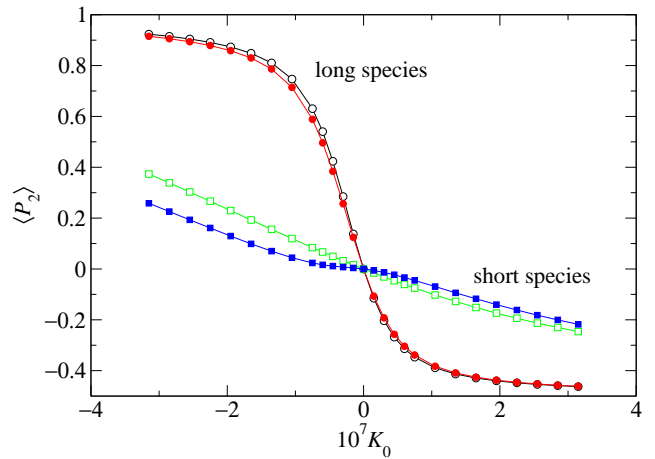}
			\caption{
				{Average nematic order parameter of rods in a bidisperse mixture of aspect ratios $L_1/\lambda=200$ and $L_2/\lambda=400$ with number fractions $x_1=x_2=0.5$ and length coupling $P=3$.  
							Open symbols show the bulk average alignment $\langle P_2\rangle$ for each species as a function of field $K_0$, while the filled symbols show the average for rods in percolating clusters only, measured in Monte Carlo simulations at the percolation threshold.} 
				}
			\label{fig:alignment}
		\end{figure}
		
	{The simulations also allow us to probe the structure of the clusters.  
	Figure \ref{fig:alignment} shows the nematic order parameter $\langle P_2\rangle$ for the same binary mixture as in Fig.~\ref{fig:simCrossing}.  
	The black and green lines (open symbols) show how the overall aligment of the rods varies with field strength.  
	These curves are independent of packing fraction due to the ideality of the rods.  
	The red and blue lines (closed symbols) show the mean alignment of rods within the percolating clusters only.}
	
	{The Figure demonstrates that rods within these clusters are, on average, more isotropically oriented than in the bulk.  
	For the longer species, the difference is slight, but for the shorter species it is quite pronounced.  
	In Fig.~\ref{fig:alignment}, the percolating clusters have been analyzed at the percolation threshold itself.  
	At higher packing fractions (deeper into the percolating regime), the percolating clusters incorporate more and more of the rods in the system and the mean alignment of rods in the clusters therefore approaches that of the bulk.  
	The difference of alignment between bulk and clusters at the percolation threshold reinforces our interpretation of how percolation occurs in these mixtures: the shorter rods act as more isotropic linkers between clusters of the more strongly aligned longer species.  
	This principle applies also for disaligning fields ($K_0>0$), where the short rods link ``layers'' of connected longer rods lying perpendicular to the field.}
		
	\section{How is universality broken?\label{sec:poly}}
	
		We have seen that the percolation threshold of sufficiently slender ideal rods in the absence of an external field depends only on the first and the second moment of the length distribution of length-polydisperse rods.
		In fact, this also turns out to be true for hard rods\,\cite{Otten2011, Nigro2013}.
		The results of the previous Section suggest, however, that the percolation threshold of ideal rods in an external quadrupole field must be a function of \textit{more} than two moments.
		As we show next, a multitude of higher order moments becomes important for weak fields, depending on the field strength and field coupling parameter.
		Before doing that, it is interesting to note that the percolation threshold of perfectly aligned and disaligned rods, in our model corresponding to infinite negative and positive field strengths, again depends on the first two moments alone.
		
		To show this for perfect alignment, we only retain the leading order term \textit{independent of the angle} between the particles in eq.\,\eqref{f}.
		Inserting this into \eqref{polyOZE} and \eqref{S} gives a percolation threshold 
		\begin{equation}
		 c_\pp (\tilde{K} \rightarrow -\infty) = \frac{1}{4\lambda} \frac{\langle L^2 \rangle}{\sqrt{\langle L^2 \rangle} + \langle L \rangle}. \label{cpparallel}
		 \end{equation}
		This percolation threshold is not only much higher than that of isotropic ideal rods, which obeys $c_\pp = 1/2$, but remains a non-trivial combination of the two moments of the length distribution.
		As a result, the percolation threshold of a mixture of rods is always lower than that of the single component in the limit of perfect alignment.
		For infinite disaligning fields with $\tilde{K}\rightarrow \infty$, the leading order term in eq.\,\eqref{f} is the one that depends on the angle between the particles.
		A similar calculation then produces $c_\pp = \pi^2 / 16$, as in the monodisperse case.		
		
		For nonzero but finite field strengths, the calculations are highly nontrivial, even if the field is weak.
		We can Taylor expand the orientational distribution function in powers of the field strength and calculate the averages $\left\langle L_i^2 P_{2l}(\cos\theta) P_{2n}(\cos\theta) \right\rangle_{i\theta}$ that enter the matrix equation \eqref{polyMatrix} up to arbitrary order.
		Inserting this into our equations, we need to expand, again, in powers of the field strength and collect terms of equal power.
		We have done this for bi-, tri- and tetradisperse rod mixtures and obtain an identical expression that, up to third order in the field strength, reads		
		\begin{align}
			c_p =&\, \frac{1}{2} + \frac{2}{405}\frac{\left\langle L^{2+P}\right\rangle^2\!\!}{\left\langle L^{2}\right\rangle^2 \! \langle L^P \rangle ^2} \tilde{K}^2 \label{expK}\\
			  - &\frac{8 \left[ 9\left\langle L^{2}\right\rangle\!\left\langle L^{2+P} \right\rangle\!\left\langle L^{2+2P}\right\rangle\! -\!  \left\langle L^{2+P}\right\rangle^3 \right]}{76545\left\langle L^{2}\right\rangle^3  \! \langle L^P \rangle ^3}  \tilde{K}^3  +  \mathcal{O}(\tilde{K}^4) . \nonumber 
		\end{align}
		As the moments of the length distribution arise only from the coefficients  $\left\langle L_i^2 P_{2l}(\cos\theta) P_{2n}(\cos\theta) \right\rangle_{i\theta}$, we conclude that eq.\,\eqref{expK} must also hold for an arbitrary number of components.
		
		\begin{figure}
			\includegraphics[width=\linewidth]{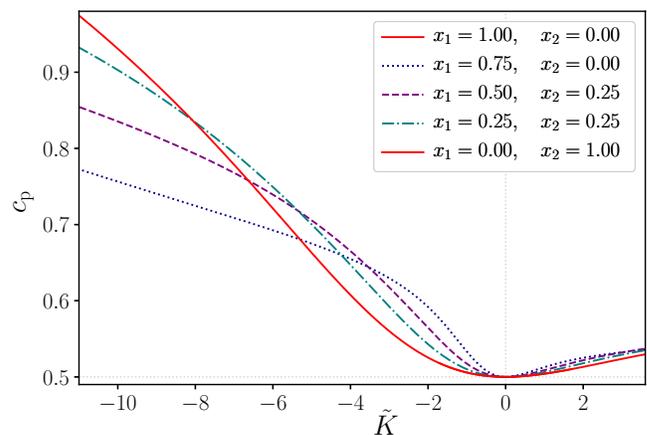}
			\caption{Dependence of the dimensionless percolation threshold $c_\pp$ on the scaled external field strength $\tilde{K}$ for a tridisperse mixture of rods with the relative aspect ratios $L_1=0.5\, L_3$ and  $L_2=0.8\, L_3$, for a length-field coupling parameter $P=3$. 
			The number fractions $x_1$ and $x_2$ are specified in the legend, and $x_3 = 1-x_1 - x_2$.}
			\label{fig:cpKwiggle}
		\end{figure}
		
		The expression \eqref{expK} shows that the percolation threshold depends on several higher moments of the length distribution whose order is determined by the field coupling parameter $P$ and the expansion order.
		This confirms that the known universal scaling of the percolation threshold with the first two moments of the length distribution fails if the field couples to the polydisperse dimensions of the rods, in agreement with Figure \ref{fig:cpKwiggle}.
		Only in the limit $P=0$ or for monodisperse systems the higher moments cancel and we recover a universal dependence of the percolation threshold $c_p$ on the scaled field strength $\tilde{K}$, as defined in eq.\,\eqref{Kwiggle}.
			
		Note that  eq.\,\eqref{expK} describes the percolation threshold for small fields very well. 
		The crossing of the curves observed in Figure \ref{fig:cpKwiggle}, however, cannot be reliably captured by the expansion, even up to order $\mathcal{O}(\tilde{K}^8)$ (results not shown).
		The reason is that the crossing takes place at effective field strengths that strongly exceed the validity range of the approximation.
		For monodisperse rods, the expansion result in eq.\,\eqref{expK} is consistent with the full solution for scaled field strengths up to $|\tilde{K}| \lesssim 4$.
		However, we observe that the agreement tends to worsen for polydisperse mixtures, where the validity range can decrease to roughly $|\tilde{K}| \lesssim 1.5$.

		\begin{figure}
			\includegraphics[width=\linewidth]{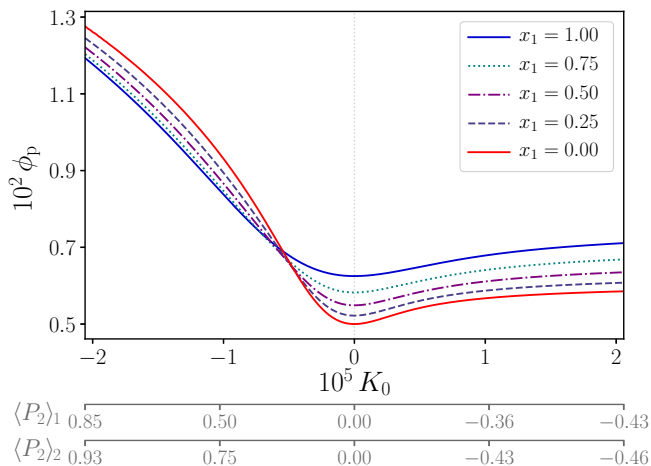}
			\caption{
				Critical volume fraction $\phi_\pp$ of a bidisperse mixture subject to a field with cubic length coupling, $P=3$, as obtained using connectedness percolation theory.
				The aspect ratios of the rods are  $L_1/\lambda = 80$ and $L_2/\lambda = 100$, with the number fractions $x_1 \in [0,1] $ and $x_2 = 1-x_1$ respectively.
				Indicated are also the nematic order parameters $\langle P_2 \rangle$ for the two types of rod.
				For strong orienting fields, the order of the curves reverses completely so that the dispersion containing only the short rods exhibits the lowest percolation threshold.}
			\label{fig:inversion}
		\end{figure}
		
\begin{figure}
			\includegraphics[width=0.98\linewidth]{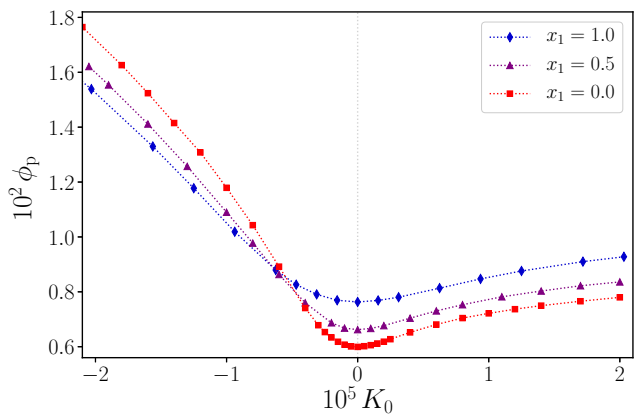}
			\caption{
				Monte Carlo simulation results for a bidisperse mixture in a field with cubic length coupling, $P=3$.
				Plotted is the critical volume fraction $\phi_\pp$ for three of the compositions shown in Figure \ref{fig:inversion}, where the aspect ratios of the rods are  $L_1/\lambda = 80$ and $L_2/\lambda = 100$.
				The field strengths $K_0$ at which the curves cross in the simulation result show excellent quantitative agreement with the crossing points in Figure \ref{fig:inversion} obtained by means of connectedness percolation theory.
				}
			\label{fig:inversion_sim}
		\end{figure}

		To investigate the crossing, we therefore need to resort to a brute-force evaluation of the full theory, vary all system parameters and hope to observe patterns.
		For this purpose, for any given ratios of the rod lengths and the coupling parameter $P$, we evaluate $c_\pp$ as a function of $K_0$ for a binary and ternary mixture, where the number fractions run from nought to unity.
		Not surprisingly, exactly because of the lack of universality, we have not been able to spot any clear trends.
		For weak fields, the percolation threshold $\phi_\pp = c_p \langle L \rangle \lambda / \langle L^2 \rangle$ increases with increasing fractions of short rods, as can be deduced from equation \eqref{expK}.
		In some cases, for sufficiently negative values of $K_0$ and for small enough difference in rod length, we find that this trend completely reverses (see Figure \ref{fig:inversion} and \ref{fig:inversion_sim}) .
		This inversion only takes place if $P$ is large enough, implying a large difference in the order parameters of different species.
		Under those conditions, mixtures with the largest fraction of short rods exhibit the lowest percolation threshold.
		Interestingly, for even more negative field strengths, we find in some cases that the original trend for very weak fields is recovered.
		Obviously,  we should recover the predictions of eq.\,\eqref{cpparallel} in the limit of perfect alignment.
		
		The Figures \ref{fig:inversion} and \ref{fig:inversion_sim} show an example of the inversion, which we observe both theoretically and in our Monte Carlo simulations.
		As already alluded to, the absolute values of the critical volume fractions obtained from the simulations are shifted with respect to the theoretical results.
		Despite that, the field strengths at which the curves cross show excellent quantitative agreement, with a relative error of less than $1\%$.

	\section{Discussion and conclusions \label{sec:dis}}
	
		In this paper, we have investigated the impact of an external alignment field in combination with polydispersity on the percolation threshold.
		For this purpose we used connectedness percolation theory and Monte Carlo simulations.
		Although we are not the first to theoretically investigate this\,\cite{Chatterjee2014, Rahatekar2010}, we have attempted to provide a considerably more comprehensive treatment.
		
		Rahatekar and collaborators\,\cite{Rahatekar2010} performed Dissipative Particle Dynamics simulations on weakly repulsive dipolar rods subject to an electrical field, neglecting dipole-dipole interactions.
		(We also ignore any field-induced interactions.)
		For rods of relatively modest aspect ratio, they also found that a bidisperse mixture in an aligning field can exhibit a lower percolation threshold than either of the individual components.
		Of course, a dipole field is not quite the same as a quadrupole field, so for comparison we also investigated the impact of a dipole field of the form $U \propto K_0 (L/\lambda)^P |\cos \gamma |$.
		In order to mimic their simulations, where alignment is achieved by fixing permanent electric charges at the ends of the rods, we would need to set $P=1$.
		In our theory, however, we only find the same behaviour when choosing $P\gtrsim2$.
		We suspect that this might be due to the second virial approximation not being sufficiently accurate for the aspect ratios of 5 and 20 used in Ref.~\cite{Rahatekar2010}.
		
		In another recent theoretical approach, Chatterjee investigated the effect of alignment and length polydispersity on the percolation threshold by mapping continuum percolation onto percolation on a Bethe lattice\,\cite{Chatterjee2014}.
		For mixtures of short and long rods, he found that bidispersity can lower the percolation threshold with respect to the monodisperse case if the long rods are sufficiently aligned.
		In that work, however, the orientations of short and long rods are decoupled, the former being isotropically oriented irrespective of the degree of order of the long ones.
		This implies that the coupling to the field does not obey Boltzmann statistics, which makes comparison to experiments difficult.
		Still, our results show almost quantitative agreement for a mixture of rods with aspect ratios $L_1/\lambda = 10$ and $L_2/\lambda = 50$ if we choose the difference in the order parameters between short and long species to be large enough, i.e., if we set $P \gtrsim 2$.
		
		In contrast to the earlier works, we believe that our theory gives a more complete view on how length polydispersity and alignment impact upon the percolation threshold and demonstrates how deeply universality is broken by the external field.
		In fact, our expansion around zero field already shows that the percolation threshold depends on several higher moments of the length distribution.
		For isotropic configurations, the percolation threshold can be made universal by an appropriate rescaling of the number density. 
		This involves a volume scale that is a function of the second moment of the distribution only.
		In non-zero field, such a rescaling is impossible, as eq.\,\eqref{expK} shows.
		
		One of the consequences of this kind of universality breaking is that, depending on the length ratios of the rods, the strength of the coupling to the external alignment field and the field strength itself, we find a wide variety of behaviors.
		This includes monodisperse shorter rods exhibiting a lower percolation threshold than longer ones, mixtures of short and long ones with a lower percolation threshold than any of the pure constituents, and a non-monotonic dependence of the composition with the lowest percolation threshold on the field strength.
		Interestingly, the full theory has to be evaluated in order to be able to observe all these effects, which happen at sufficiently large field strengths. 
		The aforementioned expansion in powers of the field strength does not reproduce this, even if we go up to eighth order. (Results not shown.)
		
		These findings are supported by the results of our Monte Carlo simulations, where we eliminated finite size effects by exploiting the fact that the universal scaling of the wrapping probability function holds even in anisotropic systems.
		The simulations are restricted to modest aspect ratios firstly because of the quadratically increasing cost of the cluster analysis for systems of longer rods, and secondly because of the need to simulate two different system sizes at each combination of field strength and system composition.
		Our theory, on the other hand, becomes more accurate the larger the aspect ratio of the rods.
		Hence, the absolute values of the theoretical percolation threshold and the simulations are shifted with respect to each other.
		It is therefore remarkable that we find excellent quantitative agreement for the field strengths at which curves for different formulations cross.\\
		
		This brings us to the last two points that we wish to address.
		The first one relates to the pre-averaging approximation that we discussed in Section \ref{sec:theory}.
		Within this approximation, we find the following expansion in powers of the scaled field strength, as defined in eq.\,\eqref{Kwiggle},		
				\begin{align}
					c_{p,\text{PA}} =& \frac{\langle L^2 \rangle}{2 \langle L \rangle^2} + \frac{1}{180} \frac{\langle L^2 \rangle \langle L^{1+P} \rangle^2}{\langle L \rangle^4 \langle L^P \rangle ^2} \tilde{K}^2 + \mathcal{O}(\tilde{K}^3) \label{cpcv}.
				\end{align}
		For monodisperse systems, this expression gives the correct zero-field percolation threshold, but the first order correction already disagrees with our exact result \eqref{expK}.
		For polydisperse rods, however, it is inaccurate even to zeroth order in $\tilde{K}$, as it fails to introduce the correct moments of the length distribution.
		It seems that, while the pre-averaging approximation is appealing because of its intuitive nature, it is fundamentally wrong for polydisperse rod dispersions.
			
		Our second point pertains to our neglect of any type of interaction between the particles.
		It is known that, for hard and for weakly attractive rods in the absence of an external field, the volume fraction at percolation for length polydisperse particles obeys the same universal scaling with the inverse weight average of the particle aspect ratio\,\cite{Otten2011, Kyrylyuk2008, Dixit2016}.
		In the presence of an aligning field, hard core interactions increase the degree of alignment\,\cite{Otten2012}.
		Because of this, it seems reasonable to suggest that the universality breaking we find in this work survives if we include hard core interactions.
		How a combination of alignment, hard core interactions and polydispersity affects the percolation threshold is unknown, and we intend to pursue it in our ongoing work.\\
	
	\begin{acknowledgments}
		The research is funded by the European Union within the Horizon 2020 project under the
DiStruc Marie Sk\l{}odowska Curie innovative training network; Grant Agreement No. 641839.
	\end{acknowledgments}
\vspace*{5mm}

	\section*{Supplementary Material}
	{All data from calculations and simulations presented in this paper are available in electronic form in the supplementary material and at https://doi.org/10.15128/r2tb09j565f.}
	
%

\end{document}